\documentclass[
  twocolumn,
  aps,
  prl,
  amsmath,
  amssymb,
  superscriptaddress,
  nofootinbib,
  floatfix
]{revtex4}
\usepackage{mathtools}
\usepackage{bm}
\usepackage{dsfont}
\usepackage{graphicx}
\usepackage{subfigure}
\usepackage{pifont}
\usepackage{textcomp}
\usepackage{gensymb}
\usepackage{accents}
\usepackage{upgreek}
\usepackage{array}
\usepackage{color}

\usepackage{hyperref}
\hypersetup{colorlinks=true, citecolor=blue, urlcolor=blue, linkcolor=blue}

\makeatletter
\newcommand*{\supplementarystart}{%
  \close@column@grid%
  \clearpage%
  \onecolumngrid%
  \setcounter{enumiv}{0} 
  \setcounter{equation}{0} 
  \setcounter{figure}{0} 
  \setcounter{table}{0} 
  \setcounter{page}{1}
  \c@secnumdepth=4
  \renewcommand{\theequation}{s\arabic{equation}} 
  \renewcommand{\bibnumfmt}[1]{[s##1]} 
  \renewcommand{\@onlinecite}{s\citealp} 
  \renewcommand{\cite}[1]{{[}\onlinecite{##1}{]}}
  \renewcommand{\thefigure}{s\arabic{figure}}
  \renewcommand{\thetable}{s\Roman{table}}
  \renewcommand{\thepage}{s\arabic{page}}
}
\makeatother

\newcommand{\pa}{\partial}

\newcommand{\be}{\begin{equation}}
\newcommand{\e}{\end{equation}}
\newcommand{\beml}{\begin{subequations}}
\newcommand{\eml}{\end{subequations}}
\newcommand{\beq}{\begin{eqnarray}}
\newcommand{\eq}{\end{eqnarray}}
\newcommand{\ba}{\begin{array}}
\newcommand{\ea}{\end{array}}
\newcommand{\bpm}{\begin{pmatrix}}
\newcommand{\epm}{\end{pmatrix}}
\newcommand{\bc}{\begin{cases}}
\newcommand{\ec}{\end{cases}}
\newcommand{\lt}{\left}
\newcommand{\rt}{\right}

\newcommand{\la}{\langle}
\newcommand{\ra}{\rangle}
\newcommand{\ep}{\varepsilon}

\renewcommand{\log}{\mathop{\mathrm{ln}}\nolimits}

\DeclareMathOperator{\tr}{Tr}

\DeclareMathOperator{\sign}{sign}

\begin{document}
	
\title{Diamagnetism of metallic nanoparticles as the result of strong spin-orbit interaction}	
\author{B. Murzaliev}
\affiliation{Radboud University, Institute for Molecules and Materials, NL-6525 AJ Nijmegen, the Netherlands}

\author{M. Titov}
\affiliation{Radboud University, Institute for Molecules and Materials, NL-6525 AJ Nijmegen, the Netherlands}
\affiliation{ITMO University, Saint Petersburg 197101, Russia}	

\author{M. I. Katsnelson}
\affiliation{Radboud University, Institute for Molecules and Materials, NL-6525 AJ Nijmegen, the Netherlands}
\affiliation{Theoretical Physics and Applied Mathematics Department, Ural Federal University, Mira Str.\,19, 620002 Ekaterinburg, Russia}

\date{\today}
\begin{abstract}
The magnetic susceptibility of an ensemble of clean metallic nanoparticles is shown to change from paramagnetic to diamagnetic one with the onset of spin-orbit interaction. The effect is quantified on the basis of symmetry analysis with the help of the random matrix theory. In particular, the magnetic susceptibility is investigated as the function of symmetry breaking parameter representing magnetic flux in the crossover from symplectic to unitary and from orthogonal to unitary ensembles. Corresponding analytical and numerical results provide a qualitative explanation to the experimental data on diamagnetism of an ensemble of gold nanorods.
\end{abstract}
\maketitle
	
\section{Introduction}
The theory of diamagnetic susceptibility of electrons in a bulk metal is non-trivial even for the case of weak magnetic fields and weakly interacting electrons  \cite{Wilsonbook,Sondheimer1960,Montambaux2015}. When applied to small metallic objects, additional complications arise due to spacial quantization. The authors of Ref.~\onlinecite{Altshuler1991} have demonstrated that the additional contribution to magnetic susceptibility can be expressed through the variance of the number of particles in a grand canonical system.  According to the theory \cite{Altshuler1991,Altshuler1993} the ensemble of tiny conducting particles should demonstrate a strong paramagnetic response in contrast to a typically diamagnetic response of large metallic systems.

Magnetic susceptibility in an ensemble of mesoscopic particles has been subsequently analyzed by a number of authors \cite{vonOppen1994,Agam1994,Richter1995,Richter1996,RichterMATH1996,RichterReview1996,GurevichShapiro1997,McCann1999,Jalabert2018} who employed various semiclassical methods to build up a quantitative theory. Still, only qualitative explanation of early experiments  \cite{LEVY1993,Benoit1993} with the arrays of ultraclean GaAs nanoparticles have been reached. Moreover, later experiments with silver and golden nanorings and nanoparticles \cite{Deblock2002,Crespo2004,Yamamoto2004dia,Dutta2007,Rojo2008,Christianen2013} demonstrated large diamagnetic response, while a series of similar experiments revealed paramagnetic susceptibility \cite{Yamamoto2004,Guerrero_2008a,Guerrero2008b,Rogalev2012,Agrachev2017} (see also Refs.~\onlinecite{Christianen2013,Jalabert2018} for the review). To explain such seemingly contradicting data one needs not only additional experimental efforts but also new theoretical insights.

In Ref.~\onlinecite{Christianen2013} a qualitative explanation for the observed diamagnetic response was suggested. The explanation has linked the diamagnetic susceptibility to the onset of strong spin-orbit interaction. The latter is quite natural to expect in such heavy metal as gold, where large spin-orbit splitting of surface states at Au(111) surface (of the order of $0.1$\,eV) has also been evidenced from the direct photoemission measurements \cite{Jensen1996,Reinert2001}. However, no detailed theory of this effect was provided at that time (the change of the sign of magnetic susceptibility due to spin-orbit coupling was hypothesized just by analogy with mesoscopic magnetoresistance \cite{Larkin1980}). The aim of this work is to present a qualitative theory of the phenomenon on the basis of the random matrix approach.

Similarly to previous works we start with the well-established thermodynamic relation \cite{Altshuler1991}
\be
\la\chi_\textrm{N}\ra=\la\chi_{\mu}\ra - \frac{1}{2}\Delta_\textrm{E}\frac{\partial^{2}}{\partial B^{2}} \la (\delta N) ^{2} \ra_{\mu},
\label{Eq:Suc}
\e
where $B$ is the magnetic field, $\chi_\mu$ and $\chi_N$ denote the statistical average of magnetic susceptibility in grand-canonical and canonical ensembles, respectively, while the angular brackets stand for the additional averaging over the ensemble of nanoparticles. This expression of Eq.~(\ref{Eq:Suc}) holds to the leading order in the ratio $\Delta_\textrm{E}/\ep_\textrm{F}$, where $\Delta_\textrm{E} \sim 10$\,meV stands for the mean level spacing in a nanoparticle and $\ep_\textrm{F}\sim 10$\,eV is the Fermi energy (to be specific, we use for estimates the parameters characteristic for the experiments [\onlinecite{Christianen2013}]). The notation $\la (\delta N) ^{2} \ra_{\mu}$ refers, at zero temperature, to the variance in the number of energy levels below the chemical potential $\mu$.

Below we compute the mean susceptibility $\la\chi_\textrm{N}\ra-\la\chi_{\mu}\ra $ from interpolating random matrix ensembles \cite{Wigner51,Wigner55,Dyson62,Mehta60,MehtaBook,Beenakker} that mimic level statistics in quantum dots with and without spin-orbit interaction.

Given large spin-orbit interaction in gold, we shall be especially interested in the crossover from the ensemble of symplectic to the ensemble of unitary random matrices: the so-called GSE-GUE crossover \cite{Alt2} (G in these abbreviations means Gaussian). Such a transitional ensemble qualitatively describes the evolution of energy level distribution with increasing magnetic field in a non-interacting ballistic quantum dot with strong spin-orbit interaction. For the sake of completeness, we also study the GOE-GUE  crossover from orthogonal to unitary ensemble that similarly describes the evolution of level statistics with increasing magnetic field in the absence of spin-orbit interaction. As we have already mentioned, the latter case received a particularly detailed attention in the literature
\cite{Dupuis,Altshuler1991,Altshuler1993,vonOppen1994,Agam1994,Richter1995,Richter1996,RichterMATH1996,RichterReview1996,GurevichShapiro1997,McCann1999,Jalabert2018} that unanimously predicted paramagnetic susceptibility. Surprisingly, there have been no such studies for the case of strong spin orbit interaction.

The quantity $\la(\delta N)^{2}\ra$ is determined, in a mesoscopic system, by the statistics of energy levels on the scale of the mean level spacing.
The statistics turns out to be universal, i.\,e. independent of the microscopic details of the Hamiltonian. Such a universality has prompted the development of the random matrix theory (RMT), first, in the context of energy spectra in heavy atomic nuclei \cite{Wigner51,Wigner55}, and later in application to disordered metallic granules \cite{GorkovEliashberg,Halperin1986}.  Since then, the RMT has been widely accepted as a standard tool to address various properties of mesoscopic systems \cite{Beenakker}. The RMT has been instrumental in addressing problems of quantum chaos \cite{Stockmannbook} and in developing general symmetry classification of non-interacting disordered systems \cite{Altland}.

Below we employ two transitional ensembles of random matrices: GOE-GUE \cite{Bohigas,Efetov} and GSE-GUE \cite{Alt2}. The energy level correlations in these ensembles have been computed analytically in Refs.~\onlinecite{Pandey1983,Mehta_1983}.

\section{Model}
Let us describe first the construction of GOE-GUE transition. A Gaussian Orthogonal Ensemble (GOE) member is a real symmetric matrix $S$ of the dimension $M\times M$ parameterized by $M(M+1)/2$ independent random real elements $S_{ij}=S_{ji}$. These elements are distributed according to the Gaussian probability density $P\propto \exp(-\tr S^2/2)$ that is manifestly invariant under orthogonal transformations.

Similarly, a member of the Gaussian Unitary Ensemble (GUE) is a random Hermitian matrix $H=S+iA$ that involve additional $M(M-1)/2$ real random variables $A_{ij}=-A_{ji}$ such that the joint probability density $P\propto \exp(-\tr H^2)$ is invariant under unitary transformations. Consequently, one can define a transitional ensemble of random matrices $H= S+i \alpha A$, where the parameter $\alpha$ varies from $0$ (GOE) to $1$ (GUE). In this parameterization, the joint probability density of the variables $S_{ij}$, $A_{ij}$, yields
\be
P_{\textrm{GOE-GUE}}\propto \exp\lt[(1+\alpha^2)\tr \lt(A^2-S^2\rt)/2\rt],
\label{Eq:ProbGOEGUE}
\e
which, for intermediate values of $\alpha$, is invariant neither under orthogonal nor under unitary rotations of the matrix $H= S+i \alpha A$. The specific feature of Eq.~(\ref{Eq:ProbGOEGUE}) is that the probability density of the eigenvalues $x_n$ of $H$ is given by the Wigner semicircle $\rho(x)=\sqrt{2M-x^2}/\pi$ independent of the value of $\alpha$. Even though the statistical properties of $H$ are not universal on the scale of the semicircle, $\sqrt{2M}$, they have a large degree of universality on the scale of the mean level spacing $\Delta$. Near the center of the semicircle the latter is given by $\Delta=1/\rho(0)=\pi/\sqrt{2M}$.

A similar construction can be proposed for the GSE-GUE crossover, that is appropriate for describing a system with strong spin-orbit interaction. Here we consider a $2M\times 2M$ matrix $H=\mathcal{S}+i\alpha\,\mathcal{A}$, that can be viewed as $M\times M$ matrix of quaternions. The latter can be represented by $\mathcal{S}=\sum_{\mu=0}^{3} S^{\mu}\sigma_{\mu}$ and  $\mathcal{A}=\sum_{\mu=0}^{3} A^{\mu}\sigma_{\mu}$, where $\mathcal{S}^\mu$ are real symmetric and $\mathcal{A}^\mu$ are real antisymmetric matrices of the dimension $M\times M$, while $\sigma_\mu$ are Pauli matrices ($\sigma_0=1$). Similarly to Eq.~(\ref{Eq:ProbGOEGUE}) a joint probability density of the independent elements $\mathcal{S}^\mu_{ij}=S^\mu_{ji}$ and $\mathcal{A}^\mu_{ij}=-\mathcal{A}^\mu_{ji}$ is chosen to be
\be
P_{\textrm{GSE-GUE}}\propto \exp\lt[(1+\alpha^2)\tr \lt(\mathcal{A}^2- \mathcal{S}^2\rt)/2\rt],
\label{Eq:ProbGSEGUE}
\e
where the parameter $\alpha$ interpolates now between GSE ($\alpha=0$) and GUE ($\alpha=1$). The density of states in the ensemble of Eq.~(\ref{Eq:ProbGSEGUE}) is also given by the Wigner surmise $\rho(x)=\sqrt{4M-x^2}/\pi$ with the spectrum width $4\sqrt{M}$ for $2M$ eigenvalues.

The crossover regime in both of the ensembles corresponds to $\alpha^2 \propto 1/M$. It is, therefore, convenient to define a dimensionless parameter $\lambda=\alpha/\Delta\sqrt{2}$, where $\Delta=1/\rho(0)$ is the mean level spacing at the center of spectrum. The parameter $\lambda$ remains finite at $M$ tends to infinity.

It can be argued \cite{Beenakker} that the dimensionless parameter $\lambda$ is set by
\be
\lambda =\gamma\, \frac{\Phi}{\Phi_0}\, \sqrt{\frac{E_\textrm{Th}}{\Delta_\textrm{E}}},
\label{Eq:lambda}
\e
where $\Phi$ is the magnetic flux piercing the system, $\Phi_0=h c/e$ is the magnetic flux quantum, $\gamma$ is a non-universal geometric factor and  $E_\textrm{Th}$ is the Thouless energy \cite{Thouless1972}. In a ballistic (ultraclean) metalic grain, the latter can be taken as $E_\textrm{Th}=\hbar v_\textrm{F}/L$, where $v_\textrm{F}$ is the Fermi velocity and $L$ is the system size. With the definition of Eq.~(\ref{Eq:lambda}) we may rewrite the leading contribution to the magnetic susceptibility in Eq.~(\ref{Eq:Suc}) as
\be
\label{final}
\la \chi_N \ra = - \frac{1}{2}\chi_0 \frac{\pa^2 \la (\delta N)^{2}\ra}{\pa \lambda^2},\qquad \chi_0 = \gamma^2\,\frac{\hbar v_\textrm{F}}{\Phi^2_0}\, L^3.
\e
One can further observe that $\chi_0 =-6\gamma^2 \chi_L\,L^3$, where $\chi_L = -e^2 k_\textrm{F}/24 \pi^2 mc^2$ is nothing but the Landau diamagnetic susceptibility per unit volume , estimating the sample cross section as $L^2$. Note that in all these estimations we assume a quadratic dispersion law for the conduction electrons, $\epsilon_p = p^2/2m$.

It is worth noting, that similarly to the expression of Eq.~(\ref{Eq:Suc}) for magnetic susceptibility, one can express mesoscopic contribution to the total magnetic moment of the system as
\be
\label{final}
\mathcal{M} = -\frac{\Delta_\textrm{E}}{2}\frac{\pa  \la (\delta N)^{2} \ra}{\pa B}=-\mathcal{M}_{0} \frac{\pa \la (\delta N)^{2}\ra}{\pa \lambda},
\e
with $\mathcal{M}_{0} = \gamma\sqrt{\hbar v_{F}\Delta_{E}L^{3}}/2\Phi_0 = \gamma\mu_\textrm{B}/\sqrt{2}$ (where we have used an estimate $\Delta_\textrm{E}=4\pi^2\ep_\textrm{F}/k_\textrm{F}^3L^3$ and the expression for Bohr magneton $\mu_\textrm{B}=e\hbar/2mc$).

In order to analyze the variance  $\la (\delta N) ^{2} \ra$ numerically we define $N$ as the number of eigenstates in a certain ``energy'' strip $x \in (-X,X)$. The universal regime corresponds to $X\ll \sqrt{M}$, hence $N\ll M$. In this case the density of eigenvalues can be regarded as constant with the mean level spacing $\Delta = \pi/\sqrt{2M}$ for GOE-GUE and $\Delta = \pi/2\sqrt{M}$ for GSE-GUE.

In order to compute  $\la (\delta N) ^{2} \ra$ analytically we refer to the corresponding formulas obtained in Refs.~\onlinecite{Pandey1983,Mehta_1983} in the limit $1\ll N\ll M$. In particular, the number of particle variance in both GOE-GUE and GSE-GUE crosssover ensembles can be conveniently expressed as
\be
\la N^2 \ra = \int_{-X}^X dx \int_{-X}^X dy\, \lt[ \rho(x)\delta(x-y)+\mathcal{R}_2(x,y)\rt],
\e
where $\mathcal{R}_2(x,  y) = \la\sum_{nm}\delta (x-x_{n}) \delta(y-x_{m})\ra$ is the level correlation function. For eigenvalues near the centrum of the spectrum ($X \ll \sqrt{M}$) we may regard $\mathcal{R}_2$ to be a function of a relative distance only,
\be
\mathcal{R}_2(x,y)= R(r)/\Delta^2,\qquad r=(x-y)/\Delta,
\e
where $\Delta=1/\rho(0)$ is a constant. Then, we obtain
\be
\label{basic}
\la (\delta N)^{2}\ra = \la N\ra-2\int_{0}^{\la N\ra}dr\; \big(\la N\ra-r \big)\,Y(r),
\e
where $Y(r)=1-R(r)$ is the two-level cluster function that has been computed analytically for the transition ensembles of Eqs.~(\ref{Eq:ProbGOEGUE}, \ref{Eq:ProbGSEGUE}) in the limit $M\to \infty$ \cite{Pandey1983,Mehta_1983}.

\begin{figure}[!htb]\centering
\centerline{\includegraphics[width=0.95\linewidth]{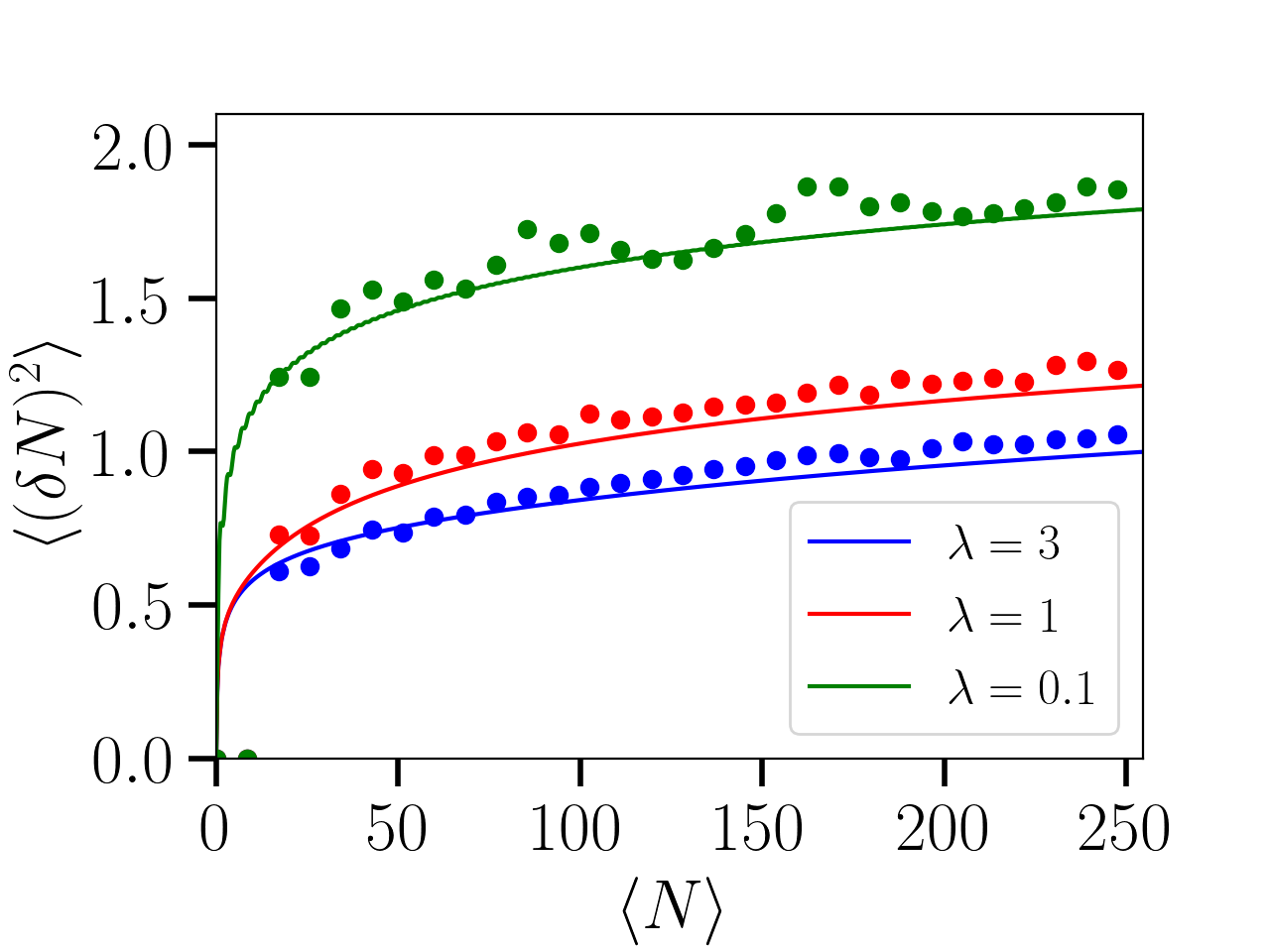}}
\caption{The variance of the number of levels $\la (\delta N)^2\ra$ versus the mean value $\la N\ra$ for the transition ensemble GSE-GOE  for different values of the crossover parameter $\lambda$ . Dots indicate the numerical data for the ensembles with $M=500$, while the corresponding solid lines are inferred from Eqs.~(\ref{basic}), (\ref{OU}), (\ref{SU}). }
\label{fig:Varmean}
\end{figure}

\begin{figure}[!htb]\centering
\includegraphics[width=0.95\linewidth]{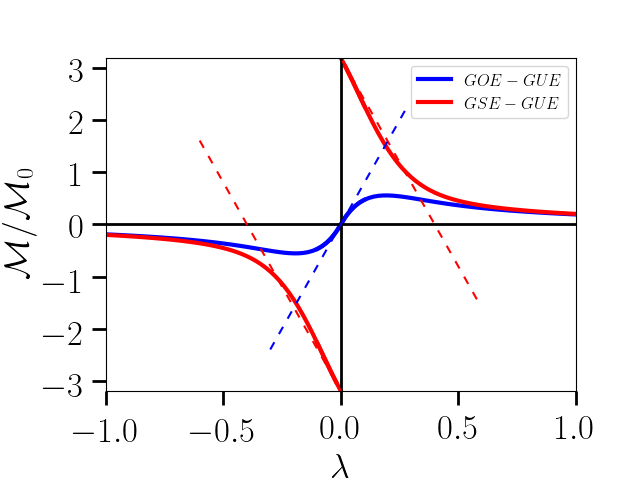}
\caption{Magnetization for the transitional ensembles obtained from Eqs.~(\ref{basic},\ref{OU},\ref{SU}): GSE-GUE (blue) and GOE-GUE (red) as a function of crossover parameter $\lambda$. Dashed lines correspond to asymptotic relations of Eq.~(\ref{assss}).}
\label{fig:magnetization}
\end{figure}

\begin{figure}[!htb]\centering
\includegraphics[width=0.95\linewidth]{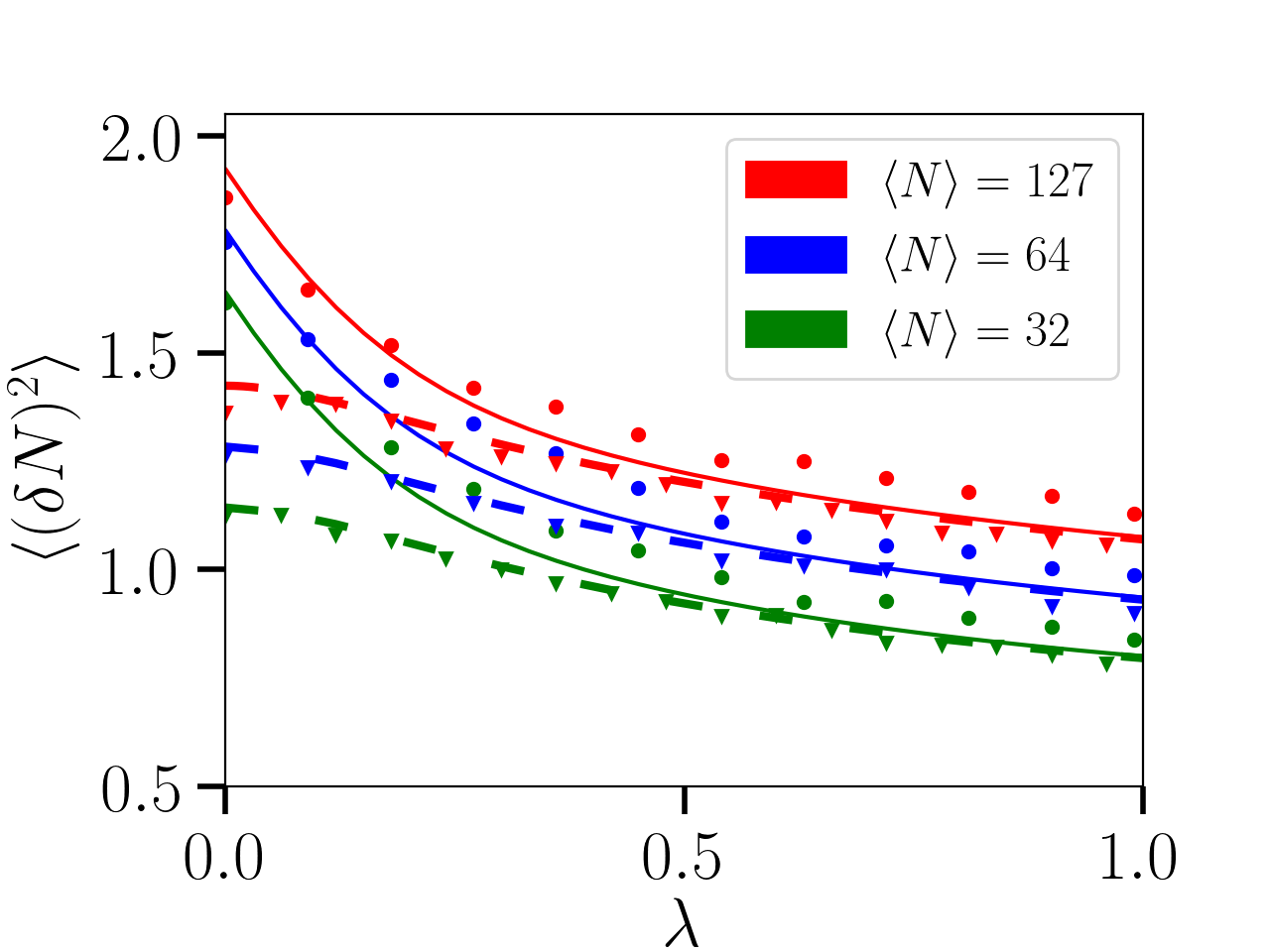}
\caption{Variance of the number of levels in a spectral region $(-X,X)$ for GSE-GUE and GOE-GUE  ensembles versus the crossover parameter $\lambda$. Solid lines and dots correspond to  analytical and numerical data of GSE-GUE respectively, while dashed lines and triangles indicate corresponding analytical results and numerical data of GOE-GUE from Eqs.~(\ref{basic},\ref{OU},\ref{SU}). }
\label{fig:variance}
\end{figure}

For a sake of completeness we reproduce here analytic results of Refs.~\onlinecite{Pandey1983,Mehta_1983} for the cluster function. For the GOE-GUE transition (\ref{Eq:ProbGOEGUE}) one finds
\beml
\label{OU}
\begin{align}
&Y_{\textrm{GOE-GUE}}(r)=\frac{\sin^2(\pi r)}{(\pi r)^2}-D(r,\lambda)J(r,\lambda),\\
&D(r,\lambda)=\frac{1}{\pi}\int_{0}^{\pi}k\,dk\,e^{2\lambda^{2}k^{2}}\sin(kr),\\
&J(r,\lambda)=\frac{1}{\pi}\int_{\pi}^{\infty}\frac{dk}{k}e^{-2\lambda^{2}k^{2}}\sin(kr),
\end{align}
\eml
while for the GSE-GUE transition (\ref{Eq:ProbGSEGUE}) one finds
\beml
\label{SU}
\begin{align}
&Y_{\textrm{GSE-GUE}}(r)=\frac{\sin^2(\pi r)}{(\pi r)^2}-I(r,\lambda)K(r,\lambda),\\
&I(r,\lambda)=\frac{1}{\pi}\int_{0}^{\pi}\frac{dk}{k}\,e^{2\lambda^{2}k^{2}}\sin kr,\\
&K(r,\lambda)=\frac{1}{\pi}\int_{\pi}^{\infty}k\,dk\,e^{-2\lambda^{2}k^{2}}\sin kr.
\end{align}
\eml	
In Figs.~\ref{fig:Varmean}, \ref{fig:magnetization}, \ref{fig:variance}, we use solid lines to plot the variance $\la (\delta N)^{2}\ra$ (obtained from Eqs.~(\ref{basic}), (\ref{OU}), (\ref{SU})) as well as its first and second derivative with respect to the parameter $\lambda$.

In Fig.~\ref{fig:Varmean} we also show the results of numerical computation of $\la (\delta N) ^{2} \ra$ from the ensemble of random matrices defined in Eqs.~(\ref{Eq:ProbGOEGUE}) and (\ref{Eq:ProbGSEGUE}) with $M=500$. Such a variance is expressed plotted there versus the mean number of states $\la N \ra=2X/\Delta$ in the strip. We find that the numerical results agree reasonably well with the results of Eqs.~(\ref{basic},\ref{OU},\ref{SU}) up to $\la N\ra \sim M/4$.

In particular, we find that the variance scales as $\ln \la N\ra$ to reproduce the well-known result \cite{Dyson62},
\be
\la (\delta N)^{2}\ra=\la N^2\ra -\la N\ra^2 = \frac{2}{\pi^{2}\beta}\log \la N\ra,
\label{Eq:altshuller}
\e
for the exact Wigner-Dyson ensembles: GOE ($\beta=1$), GUE ($\beta=2$), GSE ($\beta=4$).

It is well known, since early works of Wigner \cite{Wigner51,Wigner55}, that the mean level spacing distribution in the three Wigner-Dyson ensembles has the form
\be
P_\beta(s)=c_{\beta}\,s^{\beta}e^{-a_\beta (s\Delta/\pi)^2},
\label{Eq:LevelSpacing}
\e
where $s$ is the spacing between adjacent energy levels, $\Delta$ is the mean level spacing at the center of the semicircle,  $c_\beta$ is a normalization constant, and $a_\beta=\pi/16$, $1/\pi$, $16/9\pi$ for $\beta=1$, $2$, and $4$, respectively.  Thus, the level distribution is the most rigid (equally spaced) for GSE and the least rigid for GOE with GUE being in between. One may, therefore, naively expect that the variance $\la (\delta N)^{2}\ra$ must decrease with magnetic field in GOE-GUE crossover, but increase in GSE-GUE crossover. This logic is, however, misleading.

\begin{figure}[!htb]
\centering
\includegraphics[width=0.9\columnwidth]{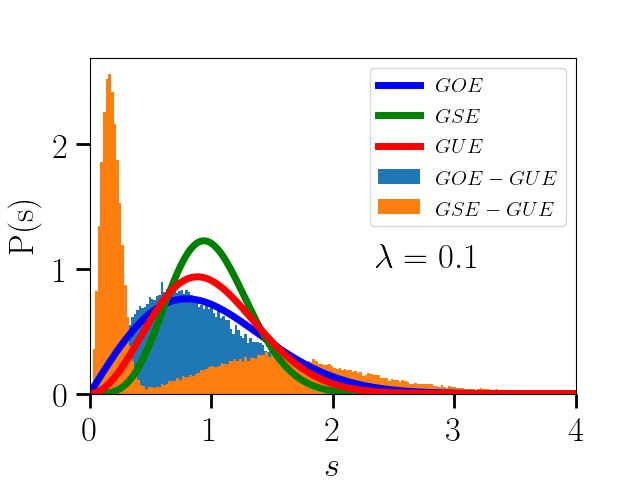}
\caption{Level spacing distribution  with $\lambda=0.1$ calculated numerically. The peak at low values of $s$ corresponds to weak lifting of Kramers degeneracy that is responsible for the leading contribution to the variance $\la (\delta N)^2\ra$. }
\label{fig:distribution}
\end{figure}

Indeed, in the absence of spin-orbit scattering, the metallic grain is characterized by $M$ double degenerate energy levels because the magnetic field, at $\Phi\sim \Phi_0$, is still too weak to induce any noticeable Zeeman splitting. Such a level degeneracy persists, therefore, even to the GUE limit. In contrast, in the GSE-GUE crossover, the Kramers degeneracy \cite{Landau1981Quantum} is completely lifted already for $\Phi\ll \Phi_0$ due to the onset of spin-orbit interaction. As the result the level distribution in the GSE-GUE crossover reveals two distinct peaks as illustrated in Fig.~\ref{fig:distribution}. The sharp peak at small spacings corresponds to weak Kramers degeneracy lifting, that provides a leading contribution to $\la (\delta N)^{2}\ra$ in the GSE-GUE crossover.

The role of the Kramers degeneracy lifting in GSE-GUE crossover is especially evident in the dependence of the total magnetic moment $\mathcal{M}$ on $\lambda$ that is illustrated in Fig.~\ref{fig:magnetization}. While, in GOE-GUE crossover, the magnetization appears to be an analytic function of $\lambda$ that vanishes in the limit of zero field, $\lambda=0$, and increases with increasing $\lambda$, in GSE-GUE crossover the behavior of $\textrm{M}$ is non-analytic at $\lambda=0$ due to the Kramers degeneracy lifting. Lifting of Kramers pairs spreads the delta function peak in the level spacing distribution (for $s=0$) that immediately results in a finite (and non-analytic) contribution to the mean magnetization.

Indeed, from Eqs.~(\ref{basic}), (\ref{OU}), (\ref{SU}) one finds asymptotic expressions for the mean mesoscopic magnetic moment
\be
\label{assss}
\mathcal{M}=8\mathcal{M}_0
\bc  \lambda+ \mathcal{O}\lt(\lambda^2\rt),&\; \textrm{GOE-GUE},\\
\frac{\sign(\lambda)}{\sqrt{2\pi}}- \lambda+ \mathcal{O}\lt(\lambda^2\rt),&\; \textrm{GSE-GUE},
\ec
\e
that clearly demonstrates a non-analytic behavior in GSE-GUE crossover. The asymptotic expressions from Eq.~(\ref{assss}) are illustrated in Fig.~\ref{fig:magnetization} with dashed lines.

As the result of this phenomenon, the variance  $\la (\delta N)^{2}\ra$ decays in both GSE-GUE and GOE-GUE transition ensembles as a functions of $\lambda$ as it is illustrated in Fig.~\ref{fig:variance}.

\begin{figure}[!htb]\centering
\includegraphics[width=0.95\linewidth]{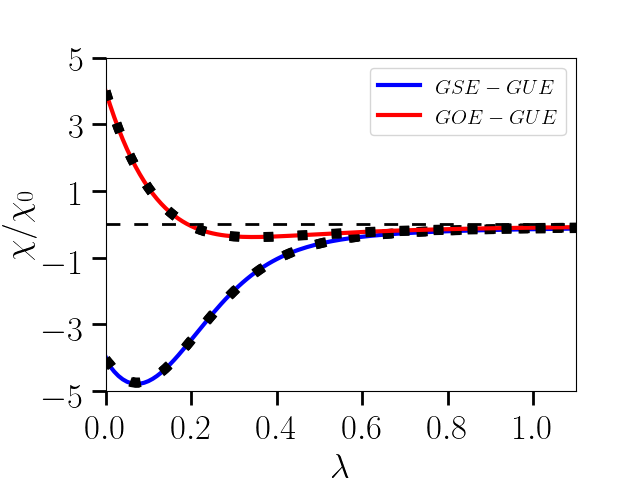}
\caption{Magnetic susceptibility for the transitional ensembles: GSE-GUE (blue) and GOE-GUE (red) as a function of crossover parameter $\lambda$. Solid lines correspond to the choice $\la N \ra =127$, while the dots do to the choice $\la N \ra = 64,32$ illustrating that the ratio $\la \chi_N\ra/\chi_0$ is independent of $\la N\ra$.}
\label{fig:magnetic}
\end{figure}

Despite $\la (\delta N)^{2}\ra$ is a monotonously decaying function of $\lambda$ for both GOE-GUE and GSE-GUE crossovers, the second derivative entering Eq.~(\ref{final}) is manifestly different in these two cases. While it changes sign from negative to positive (from paramagnetic to diamagnetic susceptibility) in GOE-GUE crossover, it stays positive (diamagnetic susceptibility) for GSE-GUE crossover (i.\,e. for systems with strong spin-orbit interactions). The behavior of the susceptibility is illustrated in Fig.~\ref{fig:magnetic} using both analytic formulas as well as numerical simulations.

Thus, the symmetry analysis performed above reaches an opposite conclusion to that expressed in Ref.~\cite{Mathur}. Namely, we find that spin-orbit interaction makes mesoscopic contribution to susceptibility entirely diamagnetic at any value of magnetic field. Instead, the susceptibility in system without spin-orbit interaction is strongly paramagnetic at small fields but becomes diamagnetic at finite fields corresponding to $\Phi/\Phi_0 \sim (\Delta_\textrm{E}/E_\textrm{Th})^{1/2}$. We note, however, that our analysis ignores the details of electron-electron interactions that may also induce a sign reversal of magnetic susceptibility \cite{Eckern1991,Principi}.

\section*{Summary}
To summarize, we computed magnetic susceptibility for metallic nanoparticles with strong spin-orbit interaction in the framework of random matrix theory. In such systems, the mesoscopic contribution to the susceptibility turns out to be entirely diamagnetic for any value of magnetic field that is consistent with recent experiments \cite{Christianen2013}. Our results, therefore, suggest that experimentally observed diamagnetism in an ensemble of metallic nanoparticles can be, indeed, a consequence of strong spin-orbit coupling. It would be interesting to check this experimentally by a systematic comparative study of light (weak spin-orbit coupling, paramagnetic response) and heavy (strong spin-orbit coupling, diamagnetic response) metals. Our consideration predicts that the enhancement factor for the diamagnetic susceptibility is independent on the size of nanoparticles but can be dependent on their geometric shape. We have also predicted a non-analytic behavior of magnetization at small fields due to Kramers degeneracy lifting in systems with strong spin-orbit interactions. The latter phenomenon can be also tested experimentally.

{\it Acknowledgments} --- The authors are thankful to Yan Fyodorov for discussions. The authors acknowledge support from the JTC-FLAGERA Project GRANSPORT and the NWO via the Spinoza Prize. M.T. acknowledges the support from the Russian Science Foundation under Project 17-12-01359.
	
\bibliographystyle{apsrev4-1}
\bibliography{mesoBIB}
	
\end{document}